\documentclass[11pt]{article}

\usepackage{amsmath}
\usepackage{amssymb}
\usepackage{amsfonts}
\usepackage{latexsym}
\usepackage{graphicx}
\usepackage{latexsym}
\usepackage{color}
\usepackage{indentfirst}
\usepackage[hyperindex]{hyperref}

\hypersetup{colorlinks=true,linkcolor=blue,citecolor=blue,urlcolor=blue}

\def\ln{\,\mbox{ln}\,}

\def\Det{\,\mbox{Det}\,}
\def\tr{\,\mbox{tr}\,}
\def\Tr{\,\mbox{Tr}\,}

\def\al{\alpha}
\def\be{\beta}
\def\ga{\gamma}
\def\Ga{\Gamma}
\def\de{\delta}
\def\De{\Delta}
\def\ep{\epsilon}

\def\la{\lambda}

\def\si{\sigma}

\def\na{\nabla}
\def\pa{\partial}

\DeclareMathOperator{\cx}{\square}

\def\beq{\begin{eqnarray}}
\def\eeq{\end{eqnarray}}
\newcommand{\eq}[1]{(\ref{#1})}
\newcommand{\n}[1]{\label{#1}}
\newcommand{\nn}{\nonumber}

\usepackage[symbol]{footmisc} 
\usepackage{float} 
\usepackage{cite}  
\usepackage{titlesec}

\titleformat*{\section}{\large\bfseries}
\titleformat*{\subsection}{\normalsize\bfseries}

\begin{document}

\begin{center}

\begin{quotation}
\begin{center}

{\Large\bf
On the total derivative divergence for a
\\
\quad
nonminimal vector operator}
\end{center}

\end{quotation}

 \vskip 7mm

\textbf{Thomas M. Sangy} 
\footnote{E-mail: thomas.sangy@estudante.ufjf.br  },
\textbf{Tib\'{e}rio de Paula Netto} 
\footnote{E-mail:  tiberio.netto@ufjf.br}
\vskip 1mm

and
\quad
\textbf{Ilya~L.~Shapiro} 
\footnote{E-mail:  ilyashapiro2003@ufjf.br}
%

\vskip 4mm

Departamento de F\'{\i}sica, \ ICE, \
Universidade Federal de Juiz de Fora - UFJF
\\
Juiz de Fora, \ 36036-900, \ MG, \ Brazil

\end{center}

\vskip 7mm

\begin{quotation}

\begin{abstract}
\vskip 2mm

\noindent
We report on the calculation of the total derivative $\Box R$ term
in the divergence of the vacuum effective action for the nonminimal
vector field operator in a curved space background. This term led to
interesting discussions in the literature, in particular because it
defines the local part of anomaly-induced effective action in
conformal quantum gravity and may be decisive for the
renormalizability of this theory. The divergent term of our interest
was previously derived several times. One of the main results is that
the mentioned local term is gauge-fixing dependent in the case of
the electromagnetic field, which contradicts the general theorems about
quantum corrections. We perform the derivation by using Riemann's
normal coordinates and confirm the previous results. The discussion
includes the possible role of the gauge-dependent IR regulators and
the related ambiguity.
\vskip 3mm

\vskip 2mm


\end{abstract}
\vskip 2mm
\end{quotation}

\section{Introduction}
\label{secIntro}

The nonminimal vector operators emerge in the gauge theories of
vector fields, either Abelian or non-Abelian, under the use of the
DeWitt-Faddeev-Popov method with the ``nonminimal'' gauge fixing
conditions. The same operators gain notorious importance for the
one-loop calculations in higher derivative quantum gravity, emerging
in the action of Faddeev-Popov ghosts and in the weight operator
\cite{frts82}. In this case, the heat-kernel representations of vector
operators contribute to the one-loop divergences, which may be of
the types $C^2$ (square of the Weyl tensor), $E_4$ (Gauss-Bonnet
topological term), $R^2$ and $\cx R$. The last term is a total
derivative and is apparently irrelevant. However, there is a big
difference between the status of this term in the general theory of
fourth-derivative gravity and in its reduced conformal sibling
\cite{NO}. In the general model, this term is irrelevant. On the
other hand, the situation may be quite different in the conformal
$C^2$-based quantum gravity, which is free from the unphysical
ghosts at the tree level.  If being multiplicatively renormalizable,
conformal quantum gravity would be also ghost-free at the quantum
level, which would be an interesting feature. The main obstacle is
that  the integration of the $\cx R$-term in the anomaly produces an
$R^2$-term in the finite one-loop contribution to the effective
action, violating conformal symmetry. If such a term appears, this
means the number of degrees of freedom of the theory changes under
the first quantum correction, the K\"all\'en-Lehmann representation
gets broken, one can expect the $R^2$-type divergences in higher
loops, such that the theory becomes non-unitary. Thus, it is
important to understand the status of $\cx R$ in the one-loop
divergences and whether this term is ambiguous or not.

The functional trace of the Schwinger-DeWitt coefficient
$\hat{a}_2$ for a nonminimal vector field operator was first
obtained in \cite{frts82} using Feynman diagrams. The result
has been confirmed as an important application of the generalized
Schwinger-DeWitt technique \cite{bavi85}. However, in both
works \cite{frts82} and \cite{bavi85} there was no evaluation of the
total derivative term $\cx R$. The first derivation of this term for the
Maxwell theory in a one-parameter gauge was given
in~\cite{Endo:1984sz}. This work concluded that the trace anomaly
for the Maxwell field depends on the gauge fixing, which led to an
alternative consideration with an opposite conclusion in
\cite{Nielsen:1988fq}. After this, for a general vector-field operator,
the trace of $\hat{a}_2$ was independently obtained in
\cite{gusynin1999} and subsequently verified in
\cite{toms2013,toms2014}. Recently, the coefficients of
the surface terms appearing in the trace of $\hat{a}_{2}$ were
independently derived in \cite{NO} using a modification of the
generalized Schwinger-DeWitt technique \cite{bavi85}, as a part
of the first calculation of the $\cx R$-divergence in conformal
quantum gravity.

All these works confirmed the original result
\cite{Endo:1984sz}. However, regardless of the methods of
\cite{Endo:1984sz}, \cite{gusynin1999}, \cite{toms2013,toms2014}
and \cite{NO} were qualitatively different, we decided to perform
one more independent derivation of the same quantity following
the approach of Refs.~\cite{bunch1979} and~\cite{panangaden1981},
i.e., using the local momentum representation and normal coordinates.

The paper is organized as follows.
The content of
Sec.~\ref{sec2} serves as a motivation for the rest of the paper.
We show the simple derivation of the desired $\cx R$-divergence
that is based on the two postulates, namely, the gauge-fixing
independence of the trace anomaly for the Maxwell field, assuming
the irrelevance of the Jacobian of the linear constant change of
variables, and on the absence of multiplicative anomaly (MA) in
the divergences of the vector operators. Both postulates may be
incorrect, but in our opinion, it is worth seeing how they can be
applied.
Sec.~\ref{sec3} describes the local momentum representation for the
propagator of the nonminimal vector operator. Sec.~\ref{sec4}
describes the coincidence limit in the propagator, the derivation of
one-loop divergences, and the potentially important gauge-fixing
dependence in the IR regulator. Finally, in
Sec.~\ref{SecConcl} we present the concluding discussions, mainly
concerning MA.

\section{Simple logic for the nonminimal vector operator}
\label{sec2}

In this section, we discuss a simple indirect derivation of the
divergences of the $\Tr \log \hat{H}$ of the nonminimal
second-order vector operator
\beq
\hat{H}
\,=\,H^{\mu}_{\nu} = \delta^{\mu}_{\nu} \cx
- \lambda \nabla^{\mu} \nabla_{\nu} + P^{\mu}_{\nu}.
\label{Hbase}
\eeq
Here $P_{\mu\nu}$ is an arbitrary tensor and $\la$ is an arbitrary
parameter. In the  case $\la=0$, we have a minimal operator, but our
interest is to elaborate the general nonminimal case with $\la\neq 0$,
and establish the $\la$-dependence in the one-loop contribution
$\,\Tr\log \hat{H}\,$ of the operator (\ref{Hbase}).

As we have already mentioned in the Introduction, this section is
based on the assumptions that
\ \textit{i)} There is no MA for the one-loop
divergences; \ \textit{ii)} There is no effect on the divergences from
the  linear constant change of variables, opposite to the statement
of \cite{Nielsen:1988fq}.
Both statements may be incorrect, but let us use these assumptions,
just to see how it works. In this section, we first formulate the
test for the correctness of $\Tr \log \hat{H}$ for the operator
(\ref{Hbase}), and then show how this test can be used in practice.

\subsection{Gauge fixing test}
\label{gaugetest}

According to the general QFT theorems, the one-loop divergences
of effective action do not depend on the gauge fixing or
parametrization of quantum fields on the classical mass shell. This
theorem was proved in many works; let us mention, in particular,
\cite{AFS,Costa-Tonin,VLT82,Kummer} for the proof of
gauge-independence on shell in the Yang-Mills theories and
\cite{Brown:1977sj} for the Maxwell theory, which is our present
case. We can also mention the proof in \cite{Lav} for the same
statement in the background field formalism, \cite{LavShap2009} for
the extension to a curved spacetime, and the proof for generic models
of quantum gravity \cite{Lavrov-renQG}. In quantum (including
conformal) gravity, this statement was applied to the separation of
essential couplings \cite{frts82,avbar86,a}. A relevant detail is that
the proof of the on-shell gauge independence in curved spacetime
\cite{LavShap2009} covers the surface terms in the
divergences.\footnote{One of the authors (I.Sh.) is grateful to
P.M. Lavrov for discussing this point.}

Let us consider a particular example with an Abelian vector model.
The vacuum effective action of the free electromagnetic field
is related to the path integral over $A_\mu$ with the action
\beq
S_1\,=\,-\,\frac14\int d^4x\sqrt{-g} \,F^2_{\mu\nu}
\,\,+\,\, \frac{\la-1}{2}\int d^4x\sqrt{-g}\, \chi^2,
\label{em}
\eeq
where $F_{\mu\nu}=\na_\mu A_\nu - \na_\nu A_\mu$,
$\chi=\na_\mu A^\mu$ is the covariant gauge-fixing condition
and $\la$ is an arbitrary gauge-fixing parameter. The one-loop
effective action of the theory is given by
\beq
\frac{i}{2}\,\Tr\log \hat{H} \,-\,i \Tr\log \hat{H}_{gh},
\label{1-loop}
\eeq
where $\hat{H}$ is the bilinear form of the action (\ref{em}) and
the second term is the ghost part,
\beq
\Tr\log \hat{H}_{gh} \,=\,\Tr\log \Big(
\frac{\de \chi}{\de A_\mu}\,\na_\mu \Big)
 \,=\,\Tr\log \cx,
\label{1-loop-ghost}
\eeq
which does not depend on $\la$. Therefore, our interest will be
in the first term in (\ref{1-loop}). The classical
equations of motion of the theory are
$\cx A^\mu - \na_\nu\na_\mu A^\nu= 0$. On the
other hand, the divergences that emerge in the free theory
(\ref{1-loop}) do not depend on $A^\mu$, but only on the
external metric. Therefore, according to the aforementioned
QFT theorem, the divergent part of the effective action
$\Ga(g)$ does not depend on the parameter $\la$.

The object of our interest is the bilinear form of the action
(\ref{em}),
\beq
&&
H_{\mu\nu}(\la)
\,=\,
g_{\mu\nu} \cx - \la \na_\nu\na_\mu  + (\la-1) R_{\mu\nu}
\,=\,
g_{\mu\nu} \cx - \la \na_\mu\na_\nu  - R_{\mu\nu}.
\label{biem}
\eeq
Compared to the general operator (\ref{Hbase}), we note that
(\ref{biem}) corresponds to the particular choice
\beq
P_{\mu\nu} \,=\, -\, R_{\mu\nu}.
\label{crit}
\eeq
As the tensor $P_{\mu\nu}$ satisfies the condition (\ref{crit}), the
dependence on the parameter $\la$ has to disappear. This condition
is satisfied for Eq.~(5.30) of~\cite{bavi85}, and it was used in this
paper as a test of correctness of all the terms except $\cx R$.

\subsection{Doubling of the  nonminimal operator} 
\label{Doubling}

The divergences of (\ref{1-loop}) with $\la=0$, have the form
\beq
\Ga^{(1)}_{div,\,min\,vec}
\,=\,-\,\frac{1}{\ep}
\int d^4x\sqrt{-g}\,\,
\Big\{ \frac{7}{60}C^2
-  \frac{8}{45}E_4
+  \frac{1}{36}\,R^2
 - \frac{1}{30}\cx R
\Big\}
\label{divminvec}
\eeq
for the vector part and
\beq
\Ga^{(1)}_{div,\,ghost}
\,=\,-\frac{1}{\ep}
\int d^4x\sqrt{-g}\,
\Big\{-\frac{1}{60}C^2
+  \frac{1}{180}E_4
-  \frac{1}{36}\,R^2
- \frac{1}{15}\cx R
\Big\}
\label{divghost}
\eeq
for the ghost contribution. Summing up, we arrive at the
well-known result
\beq
\Ga^{(1)}_{div,\,min}
\,=\,-\,\frac{1}{\ep}
\int d^4x\sqrt{-g}\,\,
\Big\{ \frac{1}{10}\,C^2
-  \frac{31}{180} E_4
 - \frac{1}{10} \cx R
\Big\},
\label{divmin}
\eeq
where $\ep=(4\pi)^2(n-4)$ is the parameter of dimensional
regularization.

In what follows, we ignore $C^2$ and $E_4$ and other
non-surface terms since they were evaluated in a consistent
way before, and concern only with the total derivative terms
\beq
\cx R,
\qquad
\na_\rho\na_\si P^{\rho\si},
\qquad
\cx P,
\qquad
\mbox{(here}
\,\,\,
P = P^{\rho\si} g_{\rho\si}
\mbox{)},
\label{surfstruct}
\eeq
for the general nonminimal operator (\ref{Hbase}). Let us note that
$2\na_\rho\na_\si R^{\rho\si} = \cx R$ owing to the Bianchi identity,
but there is no such reduction for the term
$\na_\rho\na_\si P^{\rho\si}$.

Consider the product $\hat{F} = \hat{H}(\la)\hat{H}^* (\be)$
of the general operator (\ref{Hbase}) with the
parameter $\la$ and a conjugate operator $\hat{H}^*(\be)$ of the
special form (\ref{biem}) and an arbitrary parameter $\be$,
\beq
\hat{H}^* (\be)
\,=\,
{H}_{\nu}^{*\,\mu}(\be) =
 \delta^{\mu}_{\nu} \cx
- \be \nabla^{\mu} \nabla_{\nu} - R^{\mu}_{\nu}.
\label{Hbase2}
\eeq
Assuming the absence of the MA, we get
\beq
\Tr\log \hat{F}
&=&
\Tr\log \big[ \hat{H}(\la)\,\hat{H}^*(\be)\big]
\,=\,
\Tr\log \big[H_\al^{\,\,\mu}(\la)\,{H}_{\mu}^{*\,\nu}(\be)\big]
\nn
\\
&=&
\Tr\ln \hat{H}(\la)
\,+\,
\Tr\ln \hat{H}^*(\be).
\label{prodops}
\eeq
The \textit{l.h.s.}\ of this expression can be easily evaluated if the
fourth-derivative operator $\hat{F}$ is minimal, i.e., the higher derivatives
form $\cx^2$. After a small algebra, the minimality condition for
the product (\ref{prodops}) can be found in the form
\beq
\be \,=\, \frac{\la}{\la-1}
\quad
\Longleftrightarrow
\quad
\la \,=\, \frac{\be}{\be-1}.
\label{minim}
\eeq
Taken that this is satisfied, we arrive at the minimal fourth-order
operator
\beq
\hat{F} \,=\, \hat{H}(\la)\,\hat{H}^*(\be)
&=&
\hat{1}\cx^2
\,+\,
\hat{V}^{\rho\si}\na_\rho\na_\si
\,+\,
\hat{N}^{\rho}\na_\rho
\,+\,
\hat{U},
\label{4op}
\eeq
with the following elements
\beq
&&
\hat{V}^{\rho\si}
\,=\,
[V^{\rho\si}]_\al^{\,\,\nu}
\,=\,
g^{\rho\si}\big(P_\al^{\,\,\nu} - R_\al^{\,\,\nu}\big)
\,-\,
\be \big( P_\al^{\,\,\rho} + R_\al^{\,\,\rho}\big)\,g^{\nu\si},
\nn
\\
&&
\hat{N}^\rho
\,=\,
{[\hat{N}^\rho]}_\al^{\,\,\nu}
\,=\,
-\, 2 \na^\rho R_\al^{\,\,\nu},
\nn
\\
&&
\hat{U}
\,=\,
[\hat{U}]_\al^{\,\,\nu}\,=\, - \cx R_\al^{\,\,\nu}.
\label{elemop}
\eeq
In the first of these formulas, we assume automatic symmetrization
over the pair of indices $\rho\si$, i.e., $A^\rho B^\si $ should be
replaced by $\frac12\big( A^{\rho} B^{\si} +  A^{\si} B^{\rho}\big)$.

An important detail is that, according to the general QFT theorem
discussed before, the contribution of the second term
$\Tr\log {H}_{\mu}^{\,\,\nu}(\be)$ does not depend on $\be$ and can
be derived, e.g., for $\be=0$, using Eq.~(\ref{divminvec}). Thus, all
the $\la$-dependence of $\Tr\log  \hat{F}$ is concentrated in
$\Tr\log  \hat{H}(\la)$.
Then, for the surface terms (\ref{surfstruct}) in the divergent part
of $\frac{i}{2}\Tr\ln \hat{F}$, we shall use the result of
\cite{Gusynin89}, that gives for the total derivative terms
\beq
\frac{i}{2}\Tr\ln \hat{F}\Big|_{div, \,tot-der}
&=& -\,\frac{1}{\ep}
\int d^4x\sqrt{-g}\,\tr \bigg\{
\frac{\hat{1}}{15}\cx R
\,+\,
\frac{1}{9}\cx \hat{V}
\nn
\\
&&
\qquad
-\,\,
\frac{5}{18}\na_\rho\na_\si \hat{V}^{\rho\si}
\,+\,
\frac{1}{2}\na_\rho \hat{N}^\rho
\,-\,
\hat{U}
\bigg\}.
\label{Gus4result}
\eeq
The last expression satisfies the product test
of the first calculation \cite{frts82}, i.e., Eq.~(\ref{Gus4result})
was confirmed in two independent ways and can be regarded correct.
In what follows, we obtain one more indirect confirmation of this
formula.
Using (\ref{elemop}) in Eq.~(\ref{Gus4result}), we obtain
\beq
\frac{i}{2}\Tr\ln \hat{F}\Big|_{div, \,tot-der}
&=& -\,\frac{1}{\ep}
\int d^4x\sqrt{-g}\, \bigg\{
\Big(\frac{1}{10}\,+\, \frac{\be}{36}\Big)\cx R
\nn
\\
&&
\quad
+\,\,\,\Big(\frac{1}{6}\,-\, \frac{\be}{9}\Big)\cx P
\,+\,\frac{5\be}{18}\na_\rho\na_\si P^{\rho\si}\bigg\}.
\label{op4result}
\eeq
Subtracting the contribution of the operator
$\Tr\log \hat{H}(\be)$  from~(\ref{divminvec}), we
arrive at
\beq
\frac{i}{2}\Tr\ln \hat{H}(\la)\Big|_{div, \,tot-der}
&=& -\,\frac{1}{\ep}
\int d^4x\sqrt{-g}\, \bigg\{
\Big(\frac{2}{15}\,+\, \frac{\be}{36}\Big)\cx R
\nn
\\
&&
\quad
+\,\Big(\frac{1}{6}\,-\, \frac{\be}{9}\Big)\cx P
\,+\,\frac{5\be}{18}\na_\rho\na_\si P^{\rho\si}\bigg\}.
\label{opHresult}
\eeq

Now we perform the last decisive test.
Setting $P_{\rho\si}= - R_{\rho\si}$, we should
expect that (\ref{op4result}) will not depend on $\la$. Indeed, this
works. Using (\ref{crit}), the $\be$-dependence in (\ref{op4result})
and, of course, in (\ref{opHresult}) cancels out and we end up with
\beq
\frac{i}{2}\Tr\ln \hat{H} \big(\la;\,\,P_{\rho\si}
\,=\, - \,
R_{\rho\si}\big)
\Big|_{div, \,tot-der}
&=&
-\,\frac{1}{\ep}
\int d^4x\sqrt{-g}\, \bigg\{
- \frac{1}{30}\,\cx R \bigg\},
\label{op4-min}
\eeq
independent of $\la$. Let us remark that the cancellation of
$\be$-dependence in formula (\ref{opHresult}) under the condition
(\ref{crit}) confirms both the correctness of the formula itself, but
also the correctness of the expression (\ref{Gus4result}), since the
last was used in our simple calculation.

The logic presented above looks convincing, but only until it
confronts the direct calculation. As we already mentioned in the
Introduction, the multiple calculations of the quantity of our
interest did not confirm the result (\ref{opHresult}) of our general
consideration. In the next sections, we perform one more
independent calculation and then discuss the status of the problem.

Anticipating the results of the explicit calculations presented in 
Sec.~\ref{sec4}, there is a way the arguments based on the symmetry 
may be not operational in the present case. It is well-known that the 
classical symmetry may be broken by the quantum anomaly. In the 
present case, the anomaly may has its origin in the low-energy (IR) 
divergences in the loop integrals, which are regularized by 
introducing a parameter with the non-zero mass dimension \cite{NO}.
With this parameter, the quantum vector field become massive and 
gains extra unphysical degrees of freedom. The effect of this breaks 
down the general arguments and may produce a specific type of
multiplicative anomaly, as we shall discuss below.

\section{Local momentum representation for the propagator}
\label{sec3}

Starting from this point, we derive the $\cx R$ term in the divergences
of the $\Tr \log \hat{H}$ of the nonminimal operator (\ref{Hbase}), by
using Riemann normal coordinates and the local momentum
representation. To use the dimensional regularization, all the basic
elements are defined in the $D$-dimensional space\footnote{The
contents of this and the next section is close (albeit some technical
differences) to that of Refs.~\cite{toms2013,toms2014}.}.
The Green's function $G^{\la}_{\,\mu^{\prime}} ( x, x^{\prime} )$
is the solution of the equation
\beq
\label{eq:2}
H^{\mu}_{\,\,\la} \, G^{\la}_{\,\mu^{\prime}} ( x, x^{\prime} )
\,=\, - \,\delta^{\mu}_{\,\mu^\prime} \,\de_c ( x, x^{\prime} ) ,
\eeq
where $\de_c ( x, x^{\prime} )$ is the covariant delta function,
which satisfies
\beq
\int d^D x \, \sqrt{|g(x')|} \,f(x') \,\de_c ( x, x^{\prime} ) \,=\, f(x).
\eeq
Here $|g(x)|$ is the absolute value of the metric determinant in the
point with coordinates $x\equiv x^\mu$.
It is not difficult to see that
\beq
\de_c ( x, x^{\prime} ) = |g|^{-1/4} \delta ( x - x^{\prime} ) |g'|^{-1/4},
\eeq
where $g = g(x)$, $g' = g(x')$ and $\delta ( x - x^{\prime} )$ is
the ordinary Dirac delta function. We consider
$G^{\mu}_{\mu^{\prime}} ( x, x^{\prime} )$ a bivector. This means,
for a covariant derivatives acting at the point $x$, the Green function
is a vector.

It proves helpful to eliminate the explicit metric dependence on the
{\it r.h.s} of Eq.~\eq{eq:2}. To this end, we introduce the modified
Green's function $\bar{G}^{\mu}_{\mu^{\prime}}$ through the relation
\beq
\label{eq:ModProp}
G^{\mu}_{\mu^{\prime}} ( x, x^{\prime} )
 \,=\, |g|^{-1/4} \, \bar{G}^{\mu}_{\mu^{\prime}} ( x, x^{\prime} )
 \, |g'|^{-1/4} .
\eeq
By doing so, Eq.~\eqref{eq:2} boils down to
\beq
\label{eq:eqnormal}
\bar{H}^{\mu}_{\,\la} \, \bar{G}^{\la}_{\,\mu^{\prime}}
\,=\, -\,  \delta^{\mu}_{\mu^{\prime}} \,\delta ( x-x' ) ,
\eeq
where
\beq
\label{eq:4}
\bar{H}^{\mu}_{\,\nu} \,=\,
\delta^{\mu}_{\,\nu}\, |g|^{1/4} \,\cx \, |g|^{-1/4}
\,-\, \la\, |g|^{1/4}\, \nabla^{\mu} \nabla_{\nu}\, |g|^{-1/4}
\, + \,P^{\mu}_{\,\,\nu}.
\eeq

In the local momentum representation formalism, the spacetime metric
$g_{\mu\nu}(x)$ and all related quantities are expanded near the
point $x'$ in the normal coordinates. In this point one can always
provide the metric being flat, $g_{\mu\nu} (x') = \eta_{\mu\nu}$.
Then all objects in the vicinity of this point depend on the
differences $y^\mu = x^\mu - x^{\prime\mu}$. Since our goal
is the $\cx R$-type term, we can restrict our attention to the linear
in curvature terms with up to four metric derivatives. Restricting
the equations to the corresponding order terms, the relevant
expansions have the form~\cite{petrov1969,brewin2009}
\beq
\label{g_norm}
&&
g_{\mu \nu} (x) \,=\, \eta_{\mu \nu}
- \frac{1}{3} R_{\mu \al \nu \be} \, y^{\al} y^{\be}
- \frac{1}{6} R_{\mu \al \nu \be; \ga} \, y^{\al} y^{\be} y^{\ga}
-\frac{1}{20} R_{\mu \al \nu \be ; \ga \delta} \, y^{\al} y^{\be} y^{\ga} y^{\delta}
+ \ldots,
\\
&&
g^{\mu \nu} (x) \, = \,
\eta^{\mu \nu }
+ \frac{1}{3} R^{\mu }{}_{\al }{}^{\nu }{}_{\be } \, y^{\al } y^{\be }
+ \frac{1}{6} R^{\mu }{}_{\al }{}^{\nu }{}_{\be; \ga}
\, y^{\al }  y^{\be } y^{\ga }
+ \frac{1}{20} R^{\mu }{}_{\al }{}^{\nu }{}_{\be; \ga \delta }
\, y^{\al } y^{\be } y^{\ga } y^{\delta }
+ \ldots,
\nn
\\
&&
\left| g (x) \right|^{d} \, = \,
1 - \frac{d}{3} R_{\al \be} \, y^{\al} y^{\be}
- \frac{d}{6} R_{\al \be ; \ga} \, y^{\al} y^{\be} y^{\ga}
- \frac{d}{20} R_{\al \be ; \ga \delta} \,  y^{\al} y^{\be} y^{\ga} y^{\delta}
+ \ldots,
\eeq
and also
\beq
\label{eq:GammaNorCord}
&&
\Gamma^{\la}_{\mu \nu} (x) \,=\,
- \frac{2}{3} R^{\la}{}_{\mu \nu \al} \, y^{\al}
+ \Big(\frac{1}{6} R^{\la}{}_{\al \be \mu ; \nu}
-\frac{1}{3} R^{\la}{}_{\mu \nu \al ; \be}
+ \frac{1}{12} R_{\mu \al \nu \be}{}^{; \la} \Big) y^{\al} y^{\be}
\nn
\\
&&
\qquad
+ \, \Big(
\frac{1}{40} R_{\mu \al \nu \be; \ga }{}^{; \la}
-  \frac{2}{40} R^{\la}{}_{\be \mu \al ; \nu \ga}
- \frac{2}{40} R^{\la}{}_{\be \mu \al ; \ga \nu}
- \frac{2}{20} R^{\la }{}_{\mu \nu \al ; \be \ga}
\Big) y^{\al} y^{\be} y^{\ga} + \ldots,
\\
\label{eq:PNorCord}
&&
P^{\mu}_{\nu} ( x ) \,=\,
P^{\mu}_{\nu}
\,+\, P^{\mu}_{\nu ; \al} \, y^{\al}
\,+\, \frac{1}{2} P^{\mu}{}_{\nu ; \al \be}  \, y^{\al} y^{\be}
+ \ldots
\,.
\eeq
Here and in the following, the ellipsis stands for
higher-order terms in the curvature and its covariant derivatives.
The coefficients in the expansions in $y^\al$ in the {\it r.h.s.}'s
of the above equations are evaluated at the point $x^{\prime}$.
Also, in Eq.~\eqref{eq:GammaNorCord}, the symmetrization over
$(\mu, \nu)$ is assumed, although it is not explicitly written for
the compactness of the expression.

Using Eqs.~\eq{g_norm}-\eq{eq:PNorCord} in~\eq{eq:4}, we obtain
\beq
\label{eq:HessianDec}
\bar{H}^{\mu}_{\,\,\nu}
\,=\, \bar{H}^{\mu}_{0 \,\nu} \, +\, \bar{H}^{\mu}_{1\,\nu}
\,+\, \bar{H}^{\mu}_{2 \,\nu}
\,+\, \bar{H}^{\mu}_{3 \,\nu}
\,+\, \bar{H}^{\mu}_{4\, \nu}
\,+\, \ldots ,
\eeq
where each $\bar{H}^{\mu}_{k \,\nu}$ has $k$ derivatives of the
metric. The first terms are given by
\beq
&&
\bar{H}^{\mu}_{0 \nu}
\,=\, \delta^{\mu}_{\,\nu} \pa^2   - \la \, \pa^{\mu} \pa_{\nu},
\nn
\\
&&
\label{H1}
\bar{H}^{\mu}_{1 \,\nu}  \,=\, 0,
\nn
\\
&&
\bar{H}^{\mu}_{2 \,\nu} \,= \,
P^{\mu}_{\,\nu}
+ \frac{1}{6} R \delta^{\mu}_{\,\nu}
- \frac{1}{3} \bigg( 1 - \frac{\la}{2} \bigg) R^{\mu}_{\,\nu}
+ \frac16  \big(4 R^{\mu \be}{}_{\al \nu } \partial_\be
+ 4 R^{\mu }{}_{\nu \al} {}^{ \be } \partial_{\be }
- 2 \delta^{\mu}_{\nu } R_{\,\al}^{\be } \partial_{\be }
\nn
\\
&&
\qquad
\qquad
+ \,\la  R_{\nu \al } \pa^\mu
-  \la  R^{\mu }_{\,\al } \partial_{\nu }
\big) y^{\al}
+ \frac13 \big(
\delta_{\nu }^{\mu } R_{\al \rho \be \si } \pa^\rho \pa^\si
+ \la R^{\mu }{}_{\al \be} {}^\rho  \pa_\rho\pa_\nu
\big) y^{\al} y^{\be},
\label{H2}
\eeq
\beq
&&
\bar{H}^{\mu}_{3 \,\nu}  \,=\,
\Big[
P^\mu_{\,\,\nu ; \,\al }
+  \frac{1}{6} \, \delta_{\nu }^{\mu } R_{; \al}
-  \frac{1}{2} \Big( 1 - \frac{\la}{6} \Big) R^{\mu }_{\nu ; \al }
-  \frac{1}{6}  \Big( 1 - \frac{\la}{2} \Big) R_{\nu \al }{}^{; \mu}
+ \frac{1}{2} \Big(1 + \frac{\la}{6} \Big) R^{\mu }_{\al ;\, \nu }
\Big] y^{\al}
\nn
\\
&&
\qquad\quad
+ \, \Big(
\frac{1}{3} R^{\mu }{}_{\nu \al} {}^\rho {}_{; \be} \pa_{\rho }
+ \frac{1}{3} R^{\mu \rho } {}_{ \al \nu ; \be } \partial_{\rho }
+ \frac{1}{6} R^{\mu }{}_{\al \be} {}^\rho {}_{; \nu } \pa_{\rho }
+ \frac{1}{6} R^{\mu }{}_{\al \be \nu} {}^{; \rho } \pa_{\rho }
+ \frac{1}{6} R_{\al \nu \be} {}^{\rho ; \mu} \partial_{\rho }
\nn
\\
&&
\qquad\quad
- \,\frac{1}{3} \delta^{\mu}_{\nu} R_{\al ; \be}^{\rho}\pa_\rho
-  \frac{\la}{12} R^{\mu }{}_{\al ; \be } \partial_{\nu }
+ \frac{\la}{12}  R_{\nu \al ; \be } \partial^{\mu }
+ \frac{\la}{24} R_{\al \be ; \nu } \partial^{\mu }
-  \frac{\la}{24} R_{\al \be}{}^{; \mu} \partial_{\nu }
\Big) \,y^{\al} y^{\be}
\nn
\\
&&
\qquad\quad
+ \,\frac16 \big(
\de^\mu_{\,\nu} R_{\al \rho \be \si ; \ga} \pa^\rho \pa^\si
+ \la R^{\mu}{}_{\al \be \rho ; \ga}
\partial^{\rho} \partial_{\nu}
\big) \, y^{\al} y^{\be} y^{\ga},
\label{H3}
\eeq
and
\begin{equation}
\begin{split}
\label{H4}
\bar{H}^{\mu}_{4 \,\nu}  =&
\left(
  \frac{1}{2} P^{\mu }_{\nu; \al \be }
- \frac{1}{4} R^{\mu }_{\nu ; \al \be }
+ \frac{\la}{40} R^{\mu }_{\nu ; \al \be }
+ \frac{3}{40}  \delta^{\mu}_{\nu } R_{; \al \be}
+ \frac{7}{20} R^{\mu }_{\al ; \nu \be }
+ \frac{\la}{20} R^{\mu }_{\al ; \nu \be }
\right.
\\
& \left.
+ \,\frac{1}{40} \delta^{\mu }_{\nu} \cx R_{\al \be }
-  \frac{1}{20} \cx R^{\mu }{}_{\al \nu \be }
-  \frac{3}{20}  R_{\nu \al ; \be }{}^{;\mu}
+ \frac{1}{20} \la R_{\nu \al ; \be }{}^{; \mu}
+ \frac{1}{40} \la R_{\al \be ; \nu}{}^{; \mu}
\right) y^{\al} y^{\be}
\\
&
+ \,\Bigl(
\frac{1}{10} R^{\mu }{}_{\rho \al \nu ; \be \ga } \pa^{\rho }
+ \frac{1}{10} R^{\mu }{}_{\nu \al \rho ; \be \ga } \pa^{\rho }
+ \frac{1}{10} R^{\mu }{}_{\al \be \nu ; \ga \rho } \pa^{\rho }
+ \frac{1}{10} R_{\al \nu \be \rho ; \ga}{}^{; \mu} \pa^{\rho }
\\
&
+ \,\frac{1}{10} R^{\mu }{}_{\al \be \rho ; \ga \nu } \pa^{\rho }
+ \frac{1}{10} \de^{\mu }_{\nu} R_{\al \be ; \ga \rho } \pa^{\rho }
- \frac{3}{20} \de^{\mu }_{\nu} R_{\al \rho ; \be \ga }  \pa^{\rho}
-  \frac{\la}{40}  R_{\al \be ; \ga}{}^{; \mu} \pa_{\nu }
\\
&
+ \frac{\la}{40}  R_{\al \nu ; \be \ga } \pa^{\mu }
-  \frac{\la}{40}  R^{\mu }_{\al; \be \ga } \pa_{\nu }
+ \frac{\la}{40} R_{\al \be; \ga \nu } \pa^{\mu }
 \Bigr) y^{\al} y^{\be} y^{\ga}
+  \frac{1}{20} \Bigl(
 \delta^{\mu }_{\nu} R_{\al \rho \be \si ; \ga \delta } \pa^{\rho }\pa^{\si }
\\
&
+ \,\la R^{\mu }{}_{\al \be \rho ; \ga \delta } \pa^{\rho }\pa_{\nu }
\Bigr) y^{\al} y^{\be} y^{\ga} y^{\delta}
+ \ldots\,\,.
\end{split}
\end{equation}

Now we can introduce the local momentum space representation
associated with the point $x'$ by making the Fourier transformation
\beq
\bar{G}^{\mu}_{\,\mu'}(x,x')
\,\,=\,\, \int \frac{d^Dk}{(2 \pi)^D} \,\, e^{iky}\, \bar{G}^{\mu}_{\,\mu'}(k),
\eeq
where we denoted $ky = \eta^{\al\be} k_{\al }y_{\be}$. Let us find
the solution of Eq.~\eq{eq:eqnormal} using an iterative procedure.
For this, we write the Green's function as a series
\beq
\bar{G}^{\mu}_{\mu'}(k)
\,=\,
\bar{G}^{\mu}_{1\mu'}(k) \,+\, \bar{G}^{\mu}_{2\mu'}(k)
\,+\, \bar{G}^{\mu}_{3\mu'}(k)
\,+\, \bar{G}^{\mu}_{4\mu'}(k) + \ldots,
\eeq
where $\bar{G}^{\mu}_{i\mu'}(k)$ ($i=0,1,2,3,4$) has $i$'s order
derivatives of the metric. These expressions satisfy the following
schematic equations:
\begin{align}
&
\hat{\bar{H}}_0 \hat{\bar{G}}_0 \,=\, - \hat{1},
\nn
\\
&
\hat{\bar{G}}_1 \,=\, \hat{\bar{G}}_0 \hat{\bar{H}}_1 \hat{\bar{G}}_0,
\nn
\\
&
\hat{\bar{G}}_2
\,=\, \hat{\bar{G}}_0 \hat{\bar{H}}_2 \hat{\bar{G}}_0
+ \hat{\bar{G}}_0 \hat{\bar{H}}_1 \hat{\bar{G}}_1,
\nn
\\
&
\hat{\bar{G}}_3 \,=\,\hat{\bar{G}}_0 \hat{\bar{H}}_3 \hat{\bar{G}}_0
+ \hat{\bar{G}}_0 \hat{\bar{H}}_2 \hat{\bar{G}}_1
+ \hat{\bar{G}}_0 \hat{\bar{H}}_1 \hat{\bar{G}}_2,
\nn
\\
&
\hat{\bar{G}}_4 = \hat{G}_0 \hat{\bar{H}}_4 \hat{\bar{G}}_0
+ \hat{G}_0 \hat{\bar{H}}_2 \hat{\bar{G}}_2
+ \hat{\bar{G}}_0 \hat{\bar{H}}_3 \hat{\bar{G}}_1
+ \hat{\bar{G}}_0 \hat{\bar{H}}_1 \hat{\bar{G}}_3,
\label{G4}
\end{align}
where the Fourier transform of~$\bar{H}^{\mu}_{i \,\nu}$ can be
directly obtained using the correspondence between the coordinate
and momentum spaces
\begin{equation}
\begin{split}
\pa_{\mu}  \longrightarrow  i k_{\mu} ,
\qquad
y^{\mu} \longrightarrow
- \frac{1}{i} \frac{\pa}{\pa k_{\mu}} \equiv
- \frac{1}{i} \, \pa^\mu.
\end{split}
\end{equation}
According to Eq.~\eq{H1}, we can use $\hat{H}_1 = 0$ and it
follows from the second of Eqs.~\eq{G4} that $\hat{G}_1 = 0$, which
greatly simplify  the system \eq{G4}. Also, among the Green's
function, only $\hat{G}_4 $ have terms that are linear in curvature
and have four derivatives of the metric. Furthermore, the structure
$\hat{G}_0 \hat{H}_2 \hat{G}_2$ is already $O(R^2)$ since both
$\hat{H}_2$ and  $\hat{G}_2$ are already linear in the curvatures.
Therefore, to obtain the first three coefficients until the desired
order (fourth derivatives of the metric, in our case), we need to
solve the equations
\beq
\label{eqG0}
&&
\bar{H}^{\mu}_{0 \,\la} ( k )\,\, \bar{G}^{\la}_{0\,\mu'} ( k )
\,=\, - \,\de^\mu_{\mu'}\,,
\\
\label{eqG2}
&&
\bar{G}^\mu_{2\,\mu'} ( k )
\,=\, \bar{G}^{\mu}_{0 \,\rho^{\prime}} ( k )
\,\bar{H}^{\rho}_{2 \,\sigma} ( k )
\,\bar{G}^{\sigma}_{0\, \mu^{\prime}} ( k )\,,
\\
&&
\bar{G}^\mu_{4\,\mu'} ( k )
\,=\, \bar{G}^{\mu}_{0 \,\rho^{\prime}} ( k )
\,\bar{H}^{\rho}_{4 \,\sigma} ( k )
\,\bar{G}^{\sigma}_{0\, \mu^{\prime}} ( k ) \,+\, \ldots,
\eeq
where
\beq
\label{eqG4}
\bar{H}^{\mu}_{0 \nu} (k)
\, = \,-\,k^{2} \delta^{\mu}_{\,\nu} + \la k^{\mu} k_{\nu},
\qquad \qquad
\eeq
\begin{equation}
\begin{split}
\bar{H}^{\mu}_{2\, \nu} ( k ) = & \,\,
P^{\mu }_{\nu }
+ \frac{1}{3} R^{\mu }_{\nu }
+ \frac{\la}{2} R^{\mu }_{\nu }
+ \frac{1}{6} \de_{\nu }^{\mu } R
- \frac{1}{3} \de_{\nu }^{\mu } R^{\al \be } k_{\al }  \pa_{\be }
+ \frac{\la}{2} k_{\nu } R^{\mu \la } \pa_{\la }
\\
&
- \frac{\la}{6}  R_{\nu }^{\la } k^{\mu} \pa_{\la }
+ \frac{2}{3} R^{\mu \al }{}_{\nu }{}^{\be } k_{\al} \pa_{\be }
+ \frac{2}{3}  R^{\mu }{}_{\nu }{}^{\al \be } k_{\al } \pa_{\be }
- \frac{\la}{3}   R^{\mu \be \al }{}_{\nu } k_{\al } \pa_{\be }
\\
&
-  \frac{\la}{3}  R^{\mu}{}_{\nu}{}^{\al \be} k_{\al }\pa_{\be }
+ \frac{1}{3} \delta_{\nu }^{\mu } R^{\la \tau \al \be }
k_{\al } k_{\la } \pa_{\be }\pa_{\tau}
-  \frac{\la}{3} R^{\mu \al \be \la } k_{\be} k_{\nu} \pa_{\al }\pa_{\la },
\end{split}
\end{equation}
\begin{equation}
\begin{split}
\bar{H}^{\mu}_{4 \,\nu} ( k ) \,=\,&
\left(
\frac{1}{4} R^{\mu }_{\beta ; \nu \alpha}
-  \frac{\la}{4} R^{\mu}_{\beta ; \nu \alpha }
-  \frac{1}{2} P^{\mu }_{\nu ; \alpha \beta }
-  \frac{1}{4} R^{\mu }_{\nu ; \alpha \beta }
-  \frac{7\la}{40} R^{\mu }_{\nu ; \alpha \beta }
-  \frac{3}{40} \delta_{\nu }^{\mu } R_{; \alpha \beta}
\right.
\\
&
\left.
+ \,
\frac{1}{10} R^{\mu }_{\alpha ; \nu \beta }
- \frac{\la}{10}  R^{\mu }_{\alpha ; \nu \beta }
- \frac{1}{40} \delta_{\nu }^{\mu } \cx R_{\alpha \beta }
- \frac{1}{20} \cx R^{\mu}_{\beta \nu \alpha }
- \frac{1}{20} R_{\nu \beta ; \alpha}{}^{; \mu}
\right.
\\
&
\left.
+ \,
\frac{3\la}{20} R_{\nu \beta; \alpha}{}^{; \mu}
- \frac{1}{10} R_{\nu \alpha; \beta}{}^{; \mu}
+ \frac{3\la}{40} R_{\alpha \beta ; \nu}{}^{; \mu}
\right) \pa^\be \pa^\al
+  \frac{1}{10} \bigg( 
R^{\mu}{}_{\delta \alpha \gamma; \nu \beta }
\\
&
\left.
- \, R^{\mu }{}_{\delta \nu \gamma ; \alpha \beta }
-   R^{\mu }{}_{\delta\alpha \gamma ; \nu \beta }
-   R^{\mu }{}_{\alpha \nu \delta ; \beta \gamma }
+  R_{\alpha \gamma \nu \delta ; \beta }{}^{; \mu}
-   R^{\mu}{}_{\nu \alpha \delta ; \beta \gamma }
+ \frac{\la}{2} R^{\mu }{}_{\delta \alpha \nu ; \beta \gamma }
\right.
\\
&
\left.
+ \,\frac{\la}{2} R^{\mu }{}_{\nu \alpha \delta ; \beta \gamma}
-  \frac{3\la}{4} g_{\nu \alpha } R^{\mu}{}_{\delta ; \beta \gamma }
+ \delta_{\nu }^{\mu }  R_{\alpha \gamma ; \beta \delta}
+ \frac{1}{2}\delta_{\nu }^{\mu } R_{\alpha \delta ; \beta \gamma }
-  \delta_{\nu }^{\mu } R_{\gamma \delta; \alpha \beta }
\right.
\\
&
\left.
-  \,\la g_{\nu \alpha } R^{\mu}_{\gamma ; \beta \delta}
+ \frac{3\la}{4} g_{\nu \alpha } R_{\gamma \delta ; \beta}{}^{; \mu}
\right) k^{\alpha} \pa^\de \pa^\ga \pa^\be
+ \frac{\la}{40} \left( R_{\beta \gamma ; \nu \alpha }
+ R_{\nu \gamma ; \alpha \beta } \right) k^{\mu}
\pa^\ga \pa^\be \pa^\al
\\
&
- \,\frac{1}{20} \left(
\delta_{\nu}^{\mu } R_{\alpha \zeta \beta \eta ; \gamma \delta}
- \lambda g_{\nu \alpha } R^{\mu }{}_{\eta \beta \zeta ; \gamma \delta}
\right) k^{\alpha} k^{\beta}  \pa^\ga \pa^\de \pa^\eta \pa^\zeta
+ \ldots.
\end{split}
\end{equation}

\section{The coincidence limit and divergences}
\label{sec4}

The Schwinger-DeWitt coefficients can be extracted from the
coincidence limit of the Green's function in the coordinate
representation. The coincidence limit of the propagator has a
heat kernel expansion
\begin{equation}
\label{eq:heatkernel}
\begin{split}
\bar{G}^{\mu}_{\,\nu} ( x,x )
\,=\,
& -\frac{i}{( 4 \pi )^{D/2}}
\left[
\frac{\Ga(1 - D/2)}{( m^2)^{1 - D/2}}\, a^{\,\,\mu}_{0 \,\nu}(x,x)
- \frac{\Ga (2 - D/2)}{(m^2)^{2 - D/2}}\, a^{\,\,\mu}_{1 \,\nu} ( x,x )
\right.
\\
&
\left.
+ \frac{\Ga(3 - D/ 2)}{(m^2)^{3 - D/2}}\,a^{\,\,\mu}_{2\,\nu}(x,x)
+ \cdots \right].
\end{split}
\end{equation}

In order to regulate infrared divergences that appear in the
integrals over the momentum, let us first introduce a mass
parameter $m$ in the equation for $\hat{\bar{G}}_0$,
\beq
\label{eq:G0}
\big[(-k^2 + m^2 ) \delta^\mu_{\,\nu}
\,+\, \la k^{\mu} k_{\nu} \big]
\, \bar{G}^{\nu}_{0\,\mu^{\prime}} ( k )
 \,=\, -\,\delta^\mu_{\,\mu^{\prime}}.
\eeq
It is not difficult to verify that the solution of Eq.~\eqref{eq:G0} is
\beq
\label{G0sol}
\bar{G}^{\mu}_{0 \nu^{\prime}} ( k )
\,=\,
\frac{\delta^{\mu}_{\,\,\nu^{\prime}}}{k^2-m^2}
\,+\,
\frac{ \ga \,k^{\mu} k_{\nu^{\prime}}}{( k^2 - m^2)
\big[ k^2 - \tilde{m}^2( \la ) \big]}\,,
\eeq
where
\beq
\label{massLa}
\tilde{m} ( \la) \,=\, \frac{m}{\sqrt{1-\la}}
\qquad
\mbox{and}
\qquad
\ga \,=\, \frac{\la}{1-\la}.
\eeq
One can note that the introduction of the mass term provides a
new massive parameter $ \tilde{m}( \la )$.
In case of electromagnetic field and nonminimal gauge, this
means the presence of a gauge-fixing dependent massive parameter. This
is an important detail, as we will discuss at the end of the
consideration.

As a starting point, let us derive the coefficients
$a_{0 \,\nu}^{\,\,\mu} ( x,x )$.
At the zeroth order, taking the
coincidence limit $x\to x^{\prime}$ or, equivalently $y \to 0$, we
have
\beq
\label{eq:limitG01}
&&
\bar{G}^{\mu}_{0 \,\nu} ( x,x )
\,=\,
\int \frac{d^{D} k}{(2 \pi )^{D}} \bar{G}^{\mu}_{0 \nu} ( k )
\,=\, \int \frac{d^{D} k}{(2 \pi )^{D}} \bigg\{
\frac{\delta^{\mu}_{\,\nu}}{k^2 - m^2}
+ \ga \frac{ k^{\mu} k_{\nu}}{( k^{2} - m^{2} )
\big[ k^2 - \tilde{m}^2( \la ) \big]}\bigg\}
\nn
\\
&&
\qquad \qquad \,\,\,
\,=\,
\int \frac{d^D k}{(2\pi)^D} \frac{\de^\mu_{\,\nu}}{k^2 - m^2}
\,+\, \frac{\ga}{D}
\int_0^1 d \xi \int \,
\frac{d^{D} k}{( 2 \pi )^{D}} \,
\frac{k^{2} \de^\mu_{\,\nu}}{\big[ k^2 - \xi m^2
- (1 - \xi) \tilde{m}^2 (\la) \big]^2},
\qquad
\eeq
where we introduced a Feynman integration parameter $\xi$. Using the
Feynman prescription for integrals in Minkowski space, we can use
standard formulas for the above integrals that can be found in many
textbooks, e.g., \cite{peskin}. The result is
\beq
\label{eq:limitG0}
\bar{G}^{\mu}_{0 \nu} ( x,x )
\,=\,
-\,
\frac{i \Gamma ( 1 - D/ 2 )}{( 4 \pi )^{D/2}\, (m^2)^{1 - D/2}}
\left\{ 1+ \frac{1}{D} \big[(1-\la )^{-D/2}-1\big] \right\} \de^\mu_{\,\nu}.
\eeq
By comparing Eqs.~\eqref{eq:limitG0} and \eqref{eq:heatkernel},
we get
\beq
a^{\,\mu}_{0 \,\,\nu} ( x,x )
\,=\,
\Big\{ 1 + \frac{1}{D} \big[(1-\la )^{-D/2}-1\big] \Big\}
\delta^\mu_{\,\nu},
\eeq
in agreement with \cite{toms2014}.

In the next order, the solution of Eq.~\eq{eqG2} is
\begin{equation}
\begin{split}
\label{eq:G2exp}
\bar{G}^{\mu}_{2\, \nu^{\prime}} ( k ) = & \,\,
 \frac{P^{\mu }_{\nu^{\prime} }}{(k^{2} - m^{2} )^2}
+ \frac{\delta^{\mu }_{\nu^{\prime} } R}{6 (k^{2} - m^{2} )^2}
- \frac{\la k^{\al } k_{\nu^{\prime} } R^{\mu }_{\al }}{3 (k^{2} - m^{2} )^3}
+ \frac{\la k^{\al } k^{\mu } R_{\nu^{\prime} \al }}{3 (k^{2}-m^{2} )^3}
\\
& -  \frac{2 (\la + 2) k^{\al } k^{\be }
R^{\mu }{}_{\al \nu^{\prime} \be }}{3 (k^{2} - m^{2} )^3}
 + \frac{(3 \la + 2) R^{\mu }_{\nu^{\prime} }}{6 (k^{2} - m^{2} )^2}
 + \frac{\ga^2 k^{\al } k^{\be } k^{\mu } k_{\nu^{\prime} }
 P_{\al \be }}{(k^{2} - m^{2} )^2 [k^{2} - \tilde{m}^{2} ( \la ) ]^2}
 \\
 &
 + \frac{\ga^2 k^{\al } k^{\be } k^{\mu } k_{\nu^{\prime} }
 R_{\al \be }}{(k^{2} - m^{2} )^2 [ k^{2} - \tilde{m}^{2} ( \la ) ]^2}
 - \frac{\la^2 k^{2} k^{\be } k_{\nu^{\prime} }
 R^{\mu }_{\be }}{3 (1 - \la) (k^{2} - m^{2} )^2 [k^{2}
 - \tilde{m}^{2} ( \la ) ]^2}
 \\
 &
 - \frac{\la^2 k^{2} k^{\be } k^{\mu } R_{\nu^{\prime}
 \be }}{3 (1 - \la) (k^{2} - m^{2} )^2 [ k^{2} - \tilde{m}^2(\la) ]^2}
 + \frac{\la^2 k^{2} k^{\mu } k_{\nu^{\prime} } R}{6 (1-\la) (k^{2}
 - m^{2} )^2 [ k^{2} - \tilde{m}^{2} ( \la ) ]^2}
 \\
 &
 - \frac{2 \la^2 k^{2} k^{\be } k^{\ga } R^{\mu }{}_{\be \nu^{\prime}
 \ga }}{3 (1 - \la) (k^{2} - m^{2} )^2 [ k^{2} - \tilde{m}^{2} ( \la ) ]^2}
 - \frac{\la^2 k^{2} k^{\be } k_{\nu^{\prime} }
 R^{\mu }_{\be }}{3 (1 - \la) (k^{2} - m^{2} )^3 [ k^{2}
 - \tilde{m}^{2} ( \la ) ]}
 \\
 &
 + \frac{\la^2 k^{2} k^{\be } k^{\mu } R_{\nu^{\prime} \be }}{3 (1 - \la) (k^{2} - m^{2} )^3 [ k^{2} - \tilde{m}^{2} ( \la ) ]} - \frac{2 \la^2 k^{2} k^{\be } k^{\ga } R^{\mu }{}_{\be \nu^{\prime} \ga }}{3 (1 - \la) (k^{2} - m^{2} )^3 [ k^{2} - \tilde{m}^{2} ( \la ) ]}
        \\
        &
        + \frac{\ga k^{\al } k_{\nu^{\prime}} P^{\mu }_{\al }}{(k^{2} - m^{2} )^2 [ k^{2} - \tilde{m}^{2} ( \la ) ]} + \frac{\ga k^{\al } k^{\mu } P_{\nu^{\prime} \al }}{(k^{2} - m^{2} )^2 [ k^{2} - \tilde{m}^{2} ( \la ) ]}
        \\
        &
        +  \frac{\la (2 + \la) k^{\al } k_{\nu^{\prime} } R^{\mu }_{\al }}{3 (1 - \la) ( k^{2} - m^{2} )^2 [ k^{2} - \tilde{m}^{2} ( \la ) ]} + \frac{\la^2 k^{2} R^{\mu }_{\nu^{\prime} }}{2 (1 - \la) (k^{2} - m^{2} )^2 [ k^{2} - \tilde{m}^{2} ( \la ) ]}
        \\
        &
        + \frac{\la^2 k^{\al } k^{\mu } R_{\nu^{\prime} \al }}{3 (1 - \la) ( k^{2} - m^{2} )^2 [ k^{2} - \tilde{m}^{2} ( \la ) ]} - \frac{(\la-2) \la k^{\mu } k_{\nu^{\prime} } R}{6 (1-\la) (k^{2} - m^{2} )^2 [ k^{2} - \tilde{m}^{2} ( \la ) ]} \\
        &
        -  \frac{2 \la k^{\al } k^{\be } R^{\mu }{}_{\al \nu^{\prime} \be }}{3 (k^{2} - m^{2} )^2 [ k^{2} - \tilde{m}^{2} ( \la )]}.
 \end{split}
\end{equation}
In the particular case, when $\la = 0$ and
$P^{\mu}_{\nu^{\prime}} = -R^{\mu}_{\nu^{\prime}}$, we
recover the result of Ref.~\cite{panangaden1981},
\beq
\bar{G}^{\mu}_{2 \,\nu^{\prime}} ( k )
\,=\,
\frac{\delta^{\mu }_{\,\nu^{\prime} }\, R}{6 (k^{2} - m^{2} )^2}
\,-\,\frac{2 R^{\mu }_{\nu^{\prime}}}{3 (k^{2} - m^{2} )^2}
\,-\,
\frac{4 k^{\al } k^{\be }
R^{\mu }{}_{\al \nu^{\prime} \be }}{3 (k^{2} - m^{2} )^3}.
\label{papan81}
\eeq

Taking the coincidence limit of Eq.~\eqref{eq:G2exp} and evaluating
the momentum integrals, we obtain
\beq
a^{\mu}_{1\, \nu} ( x, x )
\,=\, b_{1} \de^{\mu}_{\,\nu} P
\,+\, b_{2} P^{\mu}_{\,\nu}
\,+\, b_{3} \de^{\mu}_{\,\nu} R
\,+\, b_{4} R^{\mu}_{\,\nu},
\eeq
where $P = P^{\rho\si} g_{\rho\si}$ and
\beq
&&
b_{1} \,=\,
\frac{1}{D (D+2) ( D -2 ) \la} \left\{ (D - 2)  \la
+ 4 + \left( 1 - \la \right)^{-D/2} \big[ D\la + 2\la - 4 \big] \right\},
\nn
\\
&&
b_{2} \,=\,
\frac{1}{D (D+2) ( D -2 ) \la}
\Big\{
\big(D - 2\big)\big(D^2 - 2\big) \la
- 4 D - 2 \left( 1 - \la \right)^{-D/2} \big[ ( D + 2 )  \la - 2 D \big] \Big\},
\nn
\\
&&
b_{3} = \frac{1}{6 D (D+2) ( D -2 ) \la} \, \Bigl\{ ( D^{3} - D^{2} - 12 ) \la + 24
\nn
\\
&&
\qquad \quad
- \,( 1 - \la )^{-D/2} [ D ( D + 2 ) \la^{2}
\,-\, \left( D^{2} + 8 D + 12 \right) \la + 24 ]  \Bigr\},
\nn
\\
&&
b_4 \,= \, \frac{1}{3 D (D+2) ( D -2 ) \la}
\Bigl\{ \big(12 + 8D - 5 D^2\big) \la  - 12 D
\nn
\\
&&
\qquad \quad
+ \big( 1 - \la \big)^{-D/2}
\big[ D ( D + 2  ) \la^{2}- ( D^{2} + 8 D + 12 ) \la + 12 D \big] \Bigr\}.
\eeq
For $\la = 0$ we get
\beq
a^{\mu}_{\,1\, \nu} (x, x)
\,=\,
P^{\mu}_{\,\nu} + \frac{1}{6}\,\delta^{\mu}_{\,\nu}\, R,
\eeq
which is the well-known result for the minimal operator
\cite{panangaden1981}.


Finally, we can evaluate the $a^{\mu}_{2\,\nu}(x,x)$ coefficient
using the coincidence limit
\beq
\label{eq:G4}
\lim_{x \to x^{\prime}}\bar{G}^{\mu}_{4 \,\nu^{\prime}}(x,x^{\prime})
\,=\,
\int \frac{d^{D} k}{( 2 \pi )^D} \, \bar{G}^{\mu}_{4 \nu^\prime} (k)
\,=\, \int \frac{d^{D} k}{(2\pi)^{D}}\,\bar{G}^{\mu}_{0\,\rho} (k)
\,\bar{H}^{\rho}_{4 \,\si} (k) \,\bar{G}^{\si}_{0 \,\nu^{\prime}} ( k )
\,+\,  \ldots.
\eeq
The explicit expression for $\bar{G}^{\mu}_{4 \,\nu^\prime} (k)$ has
been obtained from \eqref{eqG4}. Unfortunately, it is too lengthy
and will not be displayed here, but can be available upon request.
The integrals over momentum in \eqref{eq:G4} can
be dealt with in a straightforward manner by means of
\textit{Wolfram Mathematica}~\cite{Mathematica}, with the tensor
algebra package \textsc{xAct}~\cite{xAct}. The evaluation was
performed using Feynman parameters, but we have to skip technical
details owing to their size.  After cumbersome calculations, the
results coming from \eqref{eq:G4} were compared with the heat
kernel expansion \eqref{eq:heatkernel}. This yields the trace of
$\hat{a}_{2}$ as
\beq
a^{\,\,\mu}_{2 \,\mu} ( x,x )
\,=\, c_1 \cx P
+ c_2 \nabla_{\al} \nabla_{\be} P^{\al \be}
+ c_3 \cx R + \ldots,
\label{a2}
\eeq
where the coefficients are given by the expressions
\beq
c_{0} \,=\, \frac{1}{6 D ( D + 2 ) ( D - 2 ) ( D - 4 ) \la^{2}},
\eeq
\begin{equation}
\begin{split}
\frac{c_{1}}{c_{0}} \,=\, & \,\,
\big( D^4 - 5 D^3 + 2 D^2 + 32 D - 96\big) \la^{2} +192  \la -96
- (1 - \la)^{-D/2} \big[( D^3 + 6 D^{2}
\\
&
+ 8 D ) \la^{3}
- ( D^{3} + 6 D^{2} + 56 D + 96 ) \la ^{2}
+ 48 (D + 4) \la - 96 \big],
\end{split}
\end{equation}

\begin{equation}
\begin{split}
\frac{c_{2}}{2 c_{0}}  \,=\,
& \,\, (5 D^{3} - 24 D^2 + 4 D + 48 ) \la^2
+ 24 \big( D^{2} - 3 D - 2\big) \la + 48 D
\qquad \qquad \quad
\\
&
+ (1 - \la)^{-\frac{D}{2} + 1} \big[ (D^3 - 4D)\la^2+24(D + 2)\la -48 D \big] ,
\end{split}
\end{equation}

\begin{equation}
\begin{split}
\frac{5 c_{3}}{c_{0}} \,=\,
& \,\, \big(
D^{5} - 5 D^{4} + 15 D^{3} - 70 D^{2} + 104 D - 240\big) \la^2
+ 120 \big( D^2 - 3 D + 6 \big) \la
\\
&
+ 240 ( D - 2 ) - ( 1 - \la )^{1 -\frac{D}{2}} \big[
 D ( D^{3} + D^2 - 4D - 4 ) \la^{3}
\\
&
 - D \big( D^{3} + 11 D^2 + 26 D + 16 ) \la^{2}
+ 120 ( D + 2 ) \la
+ 240 ( D -2 )\big] .
\end{split}
\end{equation}

To obtain the four-dimensional version of the above results, we use
the expansion
\beq
\label{laD4}
\big(1-\la\big)^{-D/2} \,=\, \frac{1}{(1-\la)^2}
\left[1 - \frac12 \log(1 - \la) (D - 4) + {\mathcal O}
\big((D - 4)^2\big)\right].
\eeq
Then, around $D=4$ we get
\beq
&&
c_1 \,=\,
\frac{8\la^2-21\la +6}{36(\la-1)\la}+\frac{2\la-1}{6 \la^2}\log (1-\la ),
\qquad \qquad \quad
\nn
\\
&&
c_{2}  \,=\,
\frac{13 \la ^2+6 \la -24}{36 (\la -1) \la }+\frac{\la +4}{6 \la ^2}
\log ( 1 - \la ),
\qquad \qquad \quad
\nn
\\
&&
 c_{3} \,=\, - \,\frac{133 \la ^2-168 \la -60}{360\la(1 - \la)}
 -\frac{\la ^2-5 \la -2}{12 \la ^2} \log ( 1 - \la ),
\label{eq:c3}
 \eeq
in the perfect correspondence with \cite{NO} and the previous
calculations \cite{Endo:1984sz,gusynin1999,toms2013,toms2014}.
In the special case $\la = 0$ (the minimal operator), the result is
\begin{align}
c_{1} \,=\, \frac{1}{6}, \qquad c_{2} \,=\, 0, \qquad c_{3}
\,=\, \frac{2}{15},
\end{align}
as expected.
For the Maxwell theory, where $P^{\,\mu}_{\nu} = - R^{\,\mu}_{\nu}$,
we get
\beq
\label{vetor}
a^{\,\,\mu}_{2\, \mu} ( x,x )
\,=\, \Big(c_3 - \frac{c_2}{2} -c_1 \Big) \cx R \,+ \ldots
\,=\, -\frac{1}{60} \big[ 2 + 5 \log ( 1 - \la ) \big] \cx R\, + \,\ldots.
\eeq

The result (\ref{vetor}) shows an explicit dependence in the gauge
parameter $\la$, in the coefficient of the term $\cx R$. There are
a few aspects of this result to discuss. First of all, the expression
\eq{vetor} agrees with the ones obtained in the
Refs.~\cite{Endo:1984sz,toms2013,toms2014}. On the other hand,
it disagrees with the general proof of the
gauge independence for the divergences in the vacuum effective
action of Maxwell's theory given in~\cite{Brown:1977sj}, as we
discussed in Sec.~\ref{sec2}. One
of the explanations is that this dependence, first detected in
\cite{Endo:1984sz}, should cancel with the contribution of the
gauge ghosts to anomaly, as discussed in \cite{Nielsen:1988fq}.
This is a plausible option, but the linear constant change of
variables, on which the arguments of \cite{Nielsen:1988fq} are
based, produces the Jacobian proportional to the delta function.
This contribution vanishes in the dimensional regularization and
since we use the last, the mentioned cancellation seems to be
impossible.


One interesting detail concerning the result (\ref{eq:c3}) was noted
in \cite{BarKal}. Consider again the relation (\ref{prodops}) from
Sec.~\ref{sec2} and assume that $\hat{H}(\la)$ has the condition
(\ref{crit}) satisfied, i.e., $P_{\rho \si} = - R_{\rho\si}$.
For the second operator, we assume relation (\ref{minim})
to hold, as in Sec.~\ref{sec2}. Then, using~\eq{vetor},
for the first factor of the product we get
\beq
\tr\, \hat{a}_2\big[\hat{H}^*(\la)\big]
\,=\,
-\,\frac{1}{60} \big[2 + 5 \log ( 1 - \la ) \big] \cx R \,+ \,\ldots\,,
\label{1a2}
\eeq
while for the operator $\hat{H}^*(\be)$ we get
\beq
\nn
\tr\, \hat{a}_2\big[\hat{H}^*(\be)\big]  &=&
-\,\frac{1}{60} \big[2 + 5 \log ( 1 - \be ) \big] \cx R \,+ \,\ldots\,,
\\
\nn
&=& 
-\,\frac{1}{60} 
\bigg[ 2 + 5 \log \Big( 1 - \frac{\la}{\la - 1} \Big) \bigg] \cx R 
\,+\, \ldots
\nn
\\
&=& -\,\frac{1}{60} \big[ 2 - 5 \log ( 1 - \la) \big] \cx R \,+\, \ldots.
\label{2a2}
\eeq
Even thought, taken separately, $\tr\,\hat{a}_2\big[H(\la)\big]$ and
and $\tr\, \hat{a}_2\big[H^*(\be)\big]$ are $\la$-dependent, the sum
\beq
&&
\tr\, \hat{a}_2\big[H(\la)\big] \,+\,
\tr\, \hat{a}_2\big[H^*(\be)\big]
\nn
\\
&&
\qquad \qquad \qquad
\,=\, -\,\frac{1}{60}\bigg[4+5\log\Big(\frac{1-\la}{1-\la}\Big)\bigg]
\cx R \,+\, \ldots
\,=\, - \,\frac{1}{15}\, \cx R \,+\, \ldots 
\eeq
is independent of $\la$. Let us stress that this remarkable
cancellation does not mean that there is no MA because the last
requires an independent cancellation of $\la$-dependence for
both individual expressions (\ref{1a2}) and (\ref{2a2}).


Another possible source for the $\la$-dependence of the
result~\eq{vetor} for the contribution of Abelian vector field may
be as follows. The common point of the calculations performed
in Refs.~\cite{NO,Endo:1984sz,toms2013,toms2014} and also in
the consideration presented above is the introduction of a massive
parameter. In our case, the mass in~\eq{eq:G0} has been introduced
to regularize infrared divergences. However, it is known that the
$m^2 \to 0$ limit in the vacuum contributions of the electromagnetic
field in curved space is discontinuous \cite{bavi85,BuGui}.
The reason is that introducing mass we change the number of
degrees of freedom, and these degrees of freedom contribute even
in the massless limit. Looking backward, it was precisely the
inclusion of the mass that led to the appearance of the massive
gauge-dependent parameter $\tilde{m}^2 (\la)$ as introduced in
 Eq.~\eq{massLa}.

Any solution of the momentum integral with a propagator
corresponding to the mass $\tilde{m}(\la)$ has, by dimensional
arguments, an overall coefficient containing the inverse of
\beq
\big[\tilde{m}^2 (\la)\big]^{\frac{D}{2}}
\,\,=\,\, \Big(\frac{m^2}{1-\la} \Big)^{D/2}.
\label{mlam}
\eeq
This factor is, according to~\eq{laD4}, the source of the
$\log (1-\la)$-terms in the final result.

It is worth discussing an alternative way to deal with the IR problem.
Another prescription for dealing with infrared divergences would
be to introduce a massive parameter only in the Feynman integrals.
In this case, the number of active degrees of freedom does not
increase.
So, instead of introducing a massive parameter in the theory, let us
first obtain the momentum representation for the propagator, make
Wick rotation and only afterwards regularize each momentum
integral according to the rule
\beq
\n{IR}
\int \frac{ d^D k}{(2\pi)^D} \frac{k_{\mu_{1}} k_{\mu_{2}}
\ldots k_{\mu_{2p}}}{(k^2)^q} \,\,\, \longrightarrow\,\,\,
\int \frac{ d^D k}{(2\pi)^D} \frac{k_{\mu_{1}} k_{\mu_{2}}
\ldots k_{\mu_{2p}}}{(k^2 + m^2)^q}
\,.
\eeq
Then, instead of solving~\eq{eq:G0}, let us consider the equation for  $\bar{G}^{\nu}_{0\,\mu^{\prime}} ( k )$ in the form
\beq
\label{eq:G0_2}
\big[-k^2  \delta^\mu_{\,\nu}
\,+\, \la k^{\mu} k_{\nu} \big]
\,\bar{G}^{\nu}_{0\,\mu^{\prime}} ( k )
 \,=\, -\,\delta^\mu_{\,\mu^{\prime}}
\eeq
with the well-known solution
\beq
\label{G0sol2}
\bar{G}^{\mu}_{0 \,\nu^{\prime}} ( k )
\,=\, \frac{\delta^{\mu}_{\,\,\nu^{\prime}}}{k^2}
\,+\, \frac{\la}{1-\la} \frac{  \,k^{\mu} k_{\nu^{\prime}}}{k^4}\,.
\eeq
After that, we can obtain the higher-order curvature correction to
the propagator using the procedure described in Sec.~\ref{sec3},
through Eqs.~(\ref{eqG2}) and~(\ref{eqG4}). Since the technical
details are very similar to the ones in the last section, we skip the
intermediate formulas.

Thus, applying the infrared regularization scheme~\eq{IR}, and
using the integration formula
\beq
&&
\int \frac{ d^D k}{(2\pi)^D} \frac{k_{\mu_{1}} k_{\mu_{2}}
\ldots k_{\mu_{2p}}}{(k^2 + m^2)^q}
= \frac{(m^2)^{D/2-q+p}}{2^p (4\pi)^{D/2}}
\,  \frac{\Ga(q-p-D/2)}{\Ga(q)} \, \de_{{\mu_{1}}{\mu_{2}}
\ldots{\mu_{2p}}},
\eeq
\beq
\qquad \qquad \qquad \mbox{where} \qquad
\de_{{\mu_{1}}{\mu_{2}} \ldots{\mu_{2p}}}
\,=\, \de_{{\mu_{1}}{\mu_{2}}} \dots \de_{\mu_{2p-1}\mu_{2p}}
\,+\, \mbox{all permutations},
\nn
\eeq
we can compare the result for the coincidence limit of the propagator
with the heat kernel expansion~\eq{eq:heatkernel}. This procedure
yields, for the case $P_{\mu\nu} = - R_{\mu\nu}$, the $D=4$ result
\beq
a^{\,\,\mu}_{2\, \mu} ( x,x )
\,=\, - \frac{\left(11 \la^2 - 35 \la + 10\right)}{300(1-\la)} \, \cx R + \ldots,
\eeq
which still depends on $\la$, but reproduces the $(-1/30) \cx R$ result
for the minimal $\la = 0$ theory.

All in all, regardless of several independent calculations of the
$\cx R$-term in the contribution of the nonminimal vector operator,
the issue is not entirely resolved. Owing to the dependence on the
IR regulator, the result may be calculation scheme-dependent, i.e., it
might be ambiguous.

\section{Concluding discussion}
\label{SecConcl}

The output (\ref{opHresult}) of the procedure of doubling described
in Sec.~\ref{sec2}, apparently confirms its consistency and also the
correctness of the formula (\ref{Gus4result}). However, the logic of
our consideration was based on the vanishing MA, i.e., the validity
of (\ref{prodops}). The violation of this condition is not impossible,
though. In the published literature, MA was first reported in the paper
\cite{MA-0} and subsequent works (see, e.g, \cite{MA-1,zeta}
and references therein) as the nonzero difference
\beq
\De_{MA}\,=\,
\log\Det \big(\hat{A}\hat{B}\big)
\,-\,
\log\Det \hat{A}
\,-\,
\log\Det \hat{B} ,
\label{MA}
\eeq
where $\hat{A}$ and $\hat{B}$ are differential operators. The result
was obtained by using zeta-regularization \cite{zeta}. On the other
hand, the MA obtained in this way cannot be regarded a physical
effect because the difference (\ref{MA}) can always be
compensated by the change of renormalization conditions
\cite{MA-2,MA-3,MA-4,Cognola1995}. The conclusion is that the MA
is possible only in those parts of effective action that are not subject
of an infinite UV renormalization. For instance, this may concern
the non-local terms in the effective action, except the leading
logarithms since those are controlled by the UV divergences. It is
worth noting that the universality of logarithmic divergences was
established in \cite{Salam51} and has been confirmed by all known
examples since then.

Eventually, the MA has been detected in the sub-leading part of
the form factors in the contributions of fermionic operators for
the massive fields
\cite{QED-Form,Dante}. The calculations which led to this result
were based on the heat kernel solution of \cite{bavi90} and on the
more traditional evaluation of the trace of the coincidence limit of
the $\hat{a}_3$ coefficient of the Schwinger-DeWitt expansion.
Qualitatively, the conclusion of Refs.~\cite{QED-Form,Dante} was
that the MA is directly related to the absence of MA in the divergent
part of effective action, i.e., by the universality of the leading
logarithms \cite{Salam51}.

In the case of massless fields, the logarithmic divergences take
control over the leading finite part of the one-loop effective action,
and it seems there is no room for the MA. However, the situation
becomes different in the theory with local conformal symmetry.
In this case, the total derivative divergent terms correspond to
the \textit{nonconformal} local terms in the effective action, which
escape the renormalization ambiguity. Recently, the MA of this
kind has been reported for the fermionic operators with torsion
\cite{AtA} and antisymmetric tensor field \cite{AAS}.

Thus, one  cannot rule out the presence of MA in the formula
(\ref{prodops}). However, the MA, which follows from the results of
\cite{gusynin1999,NO} and was discussed in detail in \cite{BarKal},
is different from the mentioned examples. First of all, this is the
first example that is not related to the fermionic operators. It is
known that fermionic operators are somehow special in the sense
they are not Hermitian in curved space \cite{Parker}. Thus, certain
anomalies in this case may be natural. On the contrary, the
nonminimal operator (\ref{Hbase}) is Hermitian. Furthermore, this
operator does not correspond to either one of the described
types of situations when the MA was previously detected.

It seems that the only one possible way out from the contradiction
with the gauge invariance is to further explore the subtle point 
related to the regularization of the infrared divergences by 
introducing a massive parameter.
As we have
seen, this operation leads to the effective gauge-dependent mass
(\ref{mlam}), and this $\la$-dependence of the mass is the source of
the MA.
In the present article, our aim was to clearly formulate the
problem and report on the new calculation that rules out the
possibility of the technical mistakes in the known results of
\cite{gusynin1999,NO} and other works.

Our final observation 
concerns the application of the nonminimal operator to the 
derivation of the finite local $R^2$-term in the one-loop effective 
action. Let us stress that there is a critical difference between the 
status of this term in conformal semiclassical gravity and 
conformal quantum gravity. In the semiclassical case, there is a 
well-known ambiguity which is equivalent to the freedom of 
adding a finite nonconformal term to the classical action of 
vacuum. This operation does not change quantum theory and 
is usually considered trivial. One important consequence of this 
is that the problem for the electromagnetic case in a nonminimal 
gauge can be regarded irrelevant. In the quantum gravity case, 
adding finite $R^2$ term to the classical action changes the 
theory in a dramatic way. E.g., after this, the derivation of the 
$\cx R$-term in the divergences makes no sense, as discussed 
recently in \cite{NO}. Thus, in quantum gravity the interest in 
the $\cx R$-type divergence concerns only conformal version of 
the theory.

\section*{Acknowledgements}
\label{secAck}

\noindent
T.M.S. is grateful to Funda\c{c}\~{a}o de Amparo \`{a} Pesquisa
do Estado de Minas Gerais (FAPEMIG) for supporting his MSc project.
I.Sh. is grateful to CNPq (Conselho Nacional de Desenvolvimento
Cient\'{i}fico e Tecnol\'{o}gico, Brazil)  for the partial support
under the grant 305122/2023-1.


\end{document}